%% file: main.tex
\documentclass[]{spie}  %

\usepackage{amsmath,amsfonts,amssymb}
\usepackage{graphicx}
\usepackage[colorlinks=true, allcolors=blue]{hyperref}
\input{packages}

\input{macro}

\newcommand{\namemaps}{\texttt{MAPS}\xspace}
\newcommand{\namemapsdata}{\texttt{MAPS-Data}\xspace}
\newcommand{\namemapstrain}{\texttt{MAPS-Train}\xspace}
\newcommand{\namemapsinvdes}{\texttt{MAPS-InvDes}\xspace}

\newcommand{\nameneur}{\texttt{NeurOLight}\xspace}

\newcommand{\namesprint}{\texttt{SP$^2$Rint}\xspace}

\title{Democratizing Electronic-Photonic AI Systems: An Open-Source AI-Infused Cross-Layer Co-Design and Design Automation Toolflow}

\author[a]{Hongjian Zhou}
\author[a]{Ziang Yin}
\author[a]{Jiaqi Gu}
\affil[a]{School of Electrical, Computer and Energy Engineering, Arizona State University}

\authorinfo{Further author information: (Send correspondence to Jiaqi Gu)\\Jiaqi Gu: E-mail: jiaqigu@asu.edu}

\begin{document} 
\maketitle

\input{doc/0_abstract}
\input{doc/1_intro}
\input{doc/2_Arch}

\input{doc/3_Sim}

\input{doc/5_Conclusion}


\input{main.bbl}
\end{document}

%% file: packages.tex
\usepackage[utf8]{inputenc} %
\usepackage[T1]{fontenc}    %
\usepackage{pifont}
\usepackage{xcolor}         %
\usepackage{booktabs}
\usepackage{hyperref}
\hypersetup{colorlinks,
    linkcolor=red,citecolor=citecolor,urlcolor=blue}
\usepackage{algorithm}
\usepackage{graphicx}
\usepackage{threeparttable}
\usepackage{multirow}
\usepackage{amsmath}
\usepackage{bm}
\usepackage{bbm}
\usepackage{amsfonts}
\usepackage[]{algpseudocode}
\usepackage{enumitem} %
\usepackage[subrefformat=parens,labelformat=parens]{subfig}
\usepackage{colortbl}
\usepackage{geometry}

\usepackage{wrapfig}
\usepackage[algo2e, ruled, vlined, linesnumbered, noend]{algorithm2e}

%% file: macro.tex
\definecolor{citecolor}{RGB}{34,139,34}
\definecolor{mydarkblue}{rgb}{0,0.08,1}
\definecolor{mydarkgreen}{rgb}{0.02,0.6,0.02}
\definecolor{mydarkred}{rgb}{0.8,0.02,0.02}
\definecolor{mydarkorange}{rgb}{0.40,0.2,0.02}
\definecolor{mypurple}{RGB}{111,0,255}
\definecolor{myred}{rgb}{1.0,0.0,0.0}
\definecolor{mygold}{rgb}{0.75,0.6,0.12}
\definecolor{myblue}{rgb}{0,0.2,0.8}
\definecolor{mydarkgray}{rgb}{0.,0.2,0.2}

\definecolor{lightred}{RGB}{255,235,235}
\definecolor{lightgreen}{RGB}{235,255,235}
\definecolor{lightblue}{RGB}{235,235,255}
\definecolor{lightcyan}{RGB}{235,255,255}
\definecolor{lightmagenta}{RGB}{255,235,255}
\definecolor{lightyellow}{RGB}{255,255,235}

\definecolor{qxkcolor}{RGB}{215,235,255}
\definecolor{softmaxcolor}{RGB}{230,235,255}
\definecolor{probxvcolor}{RGB}{255,255,235}

\definecolor{topkcolor}{RGB}{255,235,235}
\definecolor{zecolor}{RGB}{255,255,235}
\definecolor{dynacolor}{RGB}{235,255,255}

\definecolor{reviewcolor}{RGB}{0,0,200}

\algdef{SE}[DOWHILE]{Do}{doWhile}{\algorithmicdo}[1]{\algorithmicwhile\ #1}%

\newcommand{\squishlist}{
 \begin{list}{$\bullet$}
  { \setlength{\itemsep}{0pt}
     \setlength{\parsep}{3pt}
     \setlength{\topsep}{3pt}
     \setlength{\partopsep}{0pt}
     \setlength{\leftmargin}{1.5em}
     \setlength{\labelwidth}{1em}
     \setlength{\labelsep}{0.5em} } }

\newcommand{\squishend}{
  \end{list}  }

%% file: doc/0_abstract.tex
\begin{abstract}
Photonics is becoming a cornerstone technology for high-performance AI systems and scientific computing, offering unparalleled speed, parallelism, and energy efficiency. Despite this promise, the design and deployment of electronic–photonic AI systems remain highly challenging due to a steep learning curve across multiple layers, spanning device physics, circuit design, system architecture, and AI algorithms. The absence of a mature electronic-photonic design automation (EPDA) toolchain leads to long, inefficient design cycles and limits cross-disciplinary innovation and co-evolution. In this work, we present a cross-layer co-design and automation framework aimed at democratizing photonic AI system development. We begin by introducing our architecture designs for scalable photonic edge AI and Transformer inference, followed by SimPhony, an open-source modeling tool for rapid EPIC AI system evaluation and design-space exploration. We then highlight advances in AI-enabled photonic design automation, including physical AI-based Maxwell solvers, a fabrication-aware inverse design framework, and a scalable inverse training algorithm for meta-optical neural networks, enabling a scalable EPDA stack for next-generation electronic–photonic AI systems.
\end{abstract}


%% file: doc/1_intro.tex
\section{Introduction}
\label{sec:intro}

Integrated photonics has emerged as a cornerstone technology for the next generation of high-performance AI systems~\cite{ning2024photonic, NP_Nature_ahmed, NP_Nature2021_Xu, NP_Nature2021_Feldmann,Afifi2025ASTRA} and scientific computing~\cite{tang2025optical,Najafi2025OptoAligner}. 
By leveraging the fundamental properties of light, such as massive parallelism via wavelength division multiplexing~\cite{dong2024partial, Hua2025NaturePhotonicAccelerator}, picosecond-level propagation delays~\cite{hua2025integrated}, and ultra-low energy consumption, photonic accelerators offer a promising pathway to transcend the physical limitations of traditional electronic hardware. However, despite the significant potential of Electronic-Photonic Integrated Circuits (EPICs), their widespread adoption is currently hindered by a substantial barrier to entry~\cite{zhou2025toward}.
The design and deployment of photonic AI systems involve a steep learning curve that spans multiple layers: from fundamental device physics and electromagnetic wave theory to circuit-level mixed-signal interfaces, system architectures, and high-level AI algorithms. This cross-disciplinary complexity is further compounded by the absence of a mature Electronic-Photonic Design Automation (EPDA) toolchain~\cite{bogaerts2018silicon}. Unlike the highly automated EDA industry for digital electronics, photonic design still relies heavily on slow numerical simulations, manual device layout tuning, and fragmented evaluation frameworks. Such limitations lead to inefficient, non-scalable design cycles and prevent the effective co-evolution of hardware and algorithms.

To address these challenges and democratize the development of photonic AI, we present an open-source, AI-infused cross-layer co-design and design automation framework for electronic-photonic systems. Central to this framework is \texttt{SimPhony}~\cite{yin2025simphony}, a system-level modeling framework that enables rapid exploration by bridging the gap between device parameters and system performance metrics. Beyond system modeling, we leverage AI not only as the target workload but as a powerful tool for design automation itself. We introduce a suite of physical AI-based Maxwell solvers, including \texttt{NeurOLight}~\cite{gu2022NeurOLight}, \texttt{PACE}~\cite{zhu2024Pace}, and \texttt{PIC$^2$O-Sim}~\cite{pcma2024pic2osim}, which replace time-consuming numerical simulations with ultra-fast, differentiable neural operators. These solvers are seamlessly integrated into the \texttt{MAPS}~\cite{maps} infrastructure, a modular platform that unifies data generation, surrogate training, and fabrication-aware inverse design. Finally, we present the \texttt{SP$^2$RINT}~\cite{ma2025sp2rint} algorithm, which provides a scalable pathway for training meta-optical neural networks under strict physical constraints.

The remainder of this paper is organized as follows. Section~\ref{sec:arch} reviews our recent architecture designs for scalable photonic edge AI and Transformer inference, with a focus on hardware-algorithm co-optimization. Section~\ref{sec:tools} details our progress in advanced AI-assisted photonic design automation, highlighting the development of high-speed solvers and standardized inverse design frameworks. Finally, Section~\ref{sec:discussion} concludes with an outlook on our future electronic-photonic design automation toolflow.

%% file: doc/2_Arch.tex
\section{Cross-Layer Co-Design For Efficient Photonic AI Systems}
\label{sec:arch}

As workloads scale toward Large Language Models (LLMs), hardware/algorithm co-design serves as the critical technology to support large-scale, dynamic AI workloads with sufficient hardware efficiency, scalability, and flexibility.
In this section, we emphasize how cross-layer co-optimization enables next-generation photonic AI systems for real-world cloud/edge workloads.

To support modern LLMs, specifically attention-based Transformer model architectures, photonic computing cores need fundamental innovative designs to enable dynamic tensor operations with extensive optimization on signal conversion and data movement.
Our prior architecture design, \textbf{Lightening-Transformer}~\cite{NP_zhu2024lightening}, was the first accelerator developed to support the high-throughput, dynamic optical matrix-matrix multiplications to support efficient self-attention operations. 
It replaces weight-static photonic matrix units with a novel \textit{Dynamically-operated Photonic Tensor Core} (DPTC). 
At its heart is the \textbf{Dynamically-operated Dot-product (DDot)} engine, a coherent dot-product unit that enables picosecond-level operand switching and supports full-range (signed) matrix inputs without the need for hardware duplication or multiple inference passes. 
Lightening-Transformer fully unleashes the power of optics by integrating these computing cores with \textit{photonic interconnects} for inter-core data broadcast. 
By exploiting both Wavelength Division Multiplexing (WDM) for spectral parallelism and optical broadcast for intra-core operand sharing, the cross-layer-optimized architecture achieves over a 12$\times$ latency reduction compared to prior photonic accelerators.

Complementing the high-performance computing platforms in the cloud, we extend this architecture to a new version, \textbf{TeMPO}~\cite{NP_zhang2024tempo}, to address the needs of resource-constrained edge AI, where area and energy efficiency are highly restricted. 
In TeMPO, we introduce an efficient, time-multiplexed dynamic photonic tensor core. 
At the device level, TeMPO utilizes our customized, foundry-fabricated \textit{slow-light Mach-Zehnder Modulators} (SL-MZMs) that leverage enhanced light-matter interaction to achieve a footprint an order of magnitude smaller than standard PDK elements. 
To overcome the long-standing power bottleneck of Analog-to-Digital Converters (ADCs), a circuit-level innovation is employed: \emph{hierarchical partial product accumulation}. 
By aggregating photocurrents and utilizing lightweight capacitive temporal integration in the analog domain, the required ADC sampling frequency is largely reduced by a factor of $T$ (the integration time step, e.g., 60 cycles). 
This cross-layer co-design enables a compute density of 1.2~TOPS/mm$^2$ and an energy efficiency of 22.3~TOPS/W, providing a robust solution for real-time edge tasks such as voice keyword spotting and semantic segmentation.

In addition to throughput and energy efficiency, scalable photonic AI accelerators must be resilient to hardware non-idealities.
Our recent architecture SCATTER~\cite{NP_ICCAD24_Gu} exemplifies an extreme cross-layer co-design spanning device, circuit, layout, architecture, and algorithm, where a multi-step co-optimization pipeline jointly targets power/area minimization and robustness guarantees under realistic physical constraints.
\ding{202}~Starting from the bottom of the stack, SCATTER replaces communication-oriented foundry building blocks with \emph{compute-tailored} low-power slow-light modulators, unlocking substantial baseline reductions in footprint and energy.
\ding{203}~At the physical and circuit levels, SCATTER explores \emph{circuit/weight matrix co-sparsity} that enables crosstalk-aware layout to safely densify the photonic tensor core without sacrificing robustness.
\ding{204}~At the architecture level, we introduce an on-chip \emph{in-situ light redistribution} (rerouting) and power-gating mechanism that dynamically reallocates optical power to active rows/columns, enabling high-efficiency structured sparse matrix multiplication while improving effective SNR by avoiding over-driving inactive channels.
\ding{205}~Finally, SCATTER addresses the dominant electronic overhead by upgrading conventional electrical DACs to a \emph{hybrid electronic–optical segmented DAC}, combining high resolution with low power to preserve accuracy at reduced energy.
Together, this cross-stack strategy turns performance, efficiency, and robustness into co-optimized objectives, yielding \textbf{511$\times$ area compaction and 12.4$\times$ power-efficiency improvement} while largely resolving thermal crosstalk, demonstrating a practical path toward robust, sparse, and scalable photonic AI acceleration.

%% file: doc/3_Sim.tex
\section{Advanced AI-Assisted Photonic Design Automation Framework}
\label{sec:tools}

The integration of photonics into AI hardware promises to overcome the fundamental throughput and energy-efficiency limits of electronic computing. However, the path from theoretical algorithms to deployable electronic-photonic AI systems is currently obstructed by several critical design and implementation bottlenecks. 

First, the \textbf{computational wall of physical simulation} remains the most significant barrier to rapid iteration. Traditional numerical methods, while accurate, are too slow to be used in the large-scale optimization loops required for complex AI architectures. Second, as devices serve as the fundamental building blocks that dictate overall system performance, \textbf{photonic device design} has historically relied on manual heuristics, often leading to non-optimal footprints and limited performance. Finally, the \textbf{lack of a mature Electronic-Photonic Design Automation (EPDA) toolchain} prevents the holistic co-design of algorithms and hardware.

To bridge these gaps, this section highlights our recent advances in AI-assisted design automation for PIC~\cite{Fan2025DeepLearningPhotonicCAD}, spanning from system-level modeling to device-level simulation~\cite{yin2025simphony}, and photonic-device inverse design~\cite{boson, maps, gu2022NeurOLight}, to scalable training algorithms for next-generation meta-optical systems~\cite{ma2025sp2rint}.

\input{doc/3_1_Simphony}
\input{doc/3_2_MaxwellSolvers}
\input{doc/3_3_MAPS}
\input{doc/3_4_Sprint}

%% file: doc/3_1_Simphony.tex
\subsection{Simulation for Electronic-Photonic AI Systems}

EPICs offer strong promise for next-generation high-performance AI, but realizing that promise demands coordinated advances spanning devices, mixed-signal circuits, architectures, and design automation. The hybrid nature of these systems introduces tightly coupled behaviors across the stack, making it difficult even for experts to reason about interactions and performance bottlenecks. Compounding this challenge, the community still lacks a flexible, accurate, fast, and user-friendly simulation framework for EPIC AI systems, limiting systematic exploration of new hardware ideas and end-to-end evaluation on standard benchmarks.

To bridge this gap, we propose \textbf{SimPhony}~\cite{yin2025simphony}, an open-source, cross-layer modeling and simulation framework for rapid yet realistic EPIC AI evaluation. SimPhony alleviates the simulation bottleneck by integrating device-level photonic models, circuit connectivity, and architecture-level behavior into a unified, scalable abstraction. Instead of invoking expensive full-wave solvers during early-stage exploration, SimPhony relies on \emph{compact, layout-aware photonic models} that capture key physical effects, such as insertion loss, phase errors, modulation nonlinearity, noise, and thermal sensitivity, while remaining efficient enough for large architectural sweeps and algorithm-level studies.

At the circuit and architecture layers, we model heterogeneous EPIC platforms that combine photonic compute units with electronic control logic and data converters. Device- and layout-derived metrics are propagated upward to quantify system-level costs, including \emph{optical link budgets, ADC/DAC energy, device tuning overhead, and throughput-latency trade-offs}. This enables designers to assess realistic performance envelopes for photonic accelerators and optically interconnected AI systems under practical implementation constraints.

Importantly, SimPhony can interface directly with machine learning frameworks, enabling photonic non-idealities to be injected into training and inference. This supports hardware-aware learning, where models adapt to photonic noise, limited precision, and process variability. By closing the loop between physical modeling and algorithmic evaluation, SimPhony provides a foundational infrastructure for system-algorithm co-exploration and guides both architectural decisions and photonic hardware optimization toward scalable EPIC AI systems.

%% file: doc/3_2_MaxwellSolvers.tex
\input{figtex/fig_neurolight}
\subsection{Physical AI-Based Maxwell Solvers}
\label{sec:solvers}

The advancement of photonic AI systems hinges on design automation tools that can circumvent the prohibitive computational overhead of traditional electromagnetic (EM) simulation. While high-fidelity numerical solvers, such as finite-difference frequency-domain (FDFD) and finite-difference time-domain (FDTD), provide gold-standard accuracy, they often incur significant latency, ranging from minutes to hours per device. Such delays bottleneck iterative design space exploration and hinder the progress of hardware-software co-optimization.

Recently, \emph{learning-based PDE surrogates} have emerged as a promising alternative, aiming to learn direct mappings from physical configurations to solution fields. 
Notable approaches include physics-informed neural networks (PINNs)~\cite{chen2022wavey} and operator-learning frameworks like DeepONet~\cite{lu2019deeponet} and Fourier Neural Operators (FNO)~\cite{li2020fourier}. 
Despite their potential, applying these models to practical photonic devices remains challenging due to three critical limitations: (i) poor \emph{parametric generalization} across varying wavelengths, materials, and discretizations; (ii) fidelity loss when capturing \emph{multiscale} effects in high-contrast or resonant structures; and (iii) the violation of fundamental physical constraints, such as \emph{space-time causality}, which leads to non-physical dynamics and error accumulation.
Researchers have recently explored physics-augmented Maxwell solvers designed to systematically address these obstacles, enabling \textbf{fast yet reliable} simulation for photonic devices~\cite{gu2022NeurOLight, zhu2024Pace, Su2025SuperPhysNet, chen2022wavey}.

Among those efforts, we introduce \textbf{NeurOLight}~\cite{gu2022NeurOLight}, a neural operator for parametric photonic simulation that eliminates the need for per-device retraining. By embedding \textit{wave priors} (wavelength and permittivity cues) into a unified representation (Fig.~\ref{fig:neur_b}), it generalizes robustly across Maxwell PDEs and delivers a \textbf{100-200$\times$} speedup over FDFD. Figure~\ref{fig:neur_a} highlights its superior scalability across various device sizes compared to numerical simulators. Furthermore, Fig.~\ref{fig:neur_c} demonstrates that \nameneur field predictions for multimode interference devices align closely with ground-truth results, maintaining high fidelity alongside significant acceleration.

\textbf{To bridge the accuracy gap} in complex devices where high-contrast permittivity and scattering resonances challenge standard operators, \textbf{PACE}~\cite{zhu2024Pace} is proposed to further boost the prediction fidelity via two-stage error compensation. 
This model utilizes a cross-axis factorized integral kernel for parameter-efficient long-range modeling, coupled with a two-stage cascaded refinement paradigm inspired by human learning. By progressively distilling features across stages to refine coarse predictions, PACE reduces simulation error by \textbf{73\%} while achieving up to a \textbf{577$\times$} speedup over numerical simulators.

More recently, \textbf{to address the complexities of time-domain EM modeling}, where autoregressive surrogates are often prone to instability, we propose \textbf{PIC$^2$O-Sim}~\cite{pcma2024pic2osim}. This causality-aware dynamic convolutional neural operator strictly enforces the Maxwellian \textit{light cone} constraint. By integrating permittivity-conditioned, position-adaptive dynamic convolutions to capture material-dependent propagation, PIC$^2$O-Sim enables high-fidelity spatiotemporal simulation with up to a \textbf{310$\times$} speedup over industry-standard software like Meep~\cite{oskooi2010meep}, providing a robust differentiable engine for large-scale inverse design.

%% file: figtex/fig_neurolight.tex
\begin{figure}
    \centering
    \subfloat[]{\includegraphics[width=0.22\columnwidth]{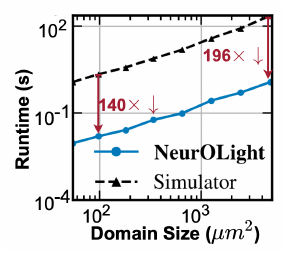}
    \label{fig:neur_a}
    }
    \hspace{20pt}
    \subfloat[]{\includegraphics[width=0.5\columnwidth]{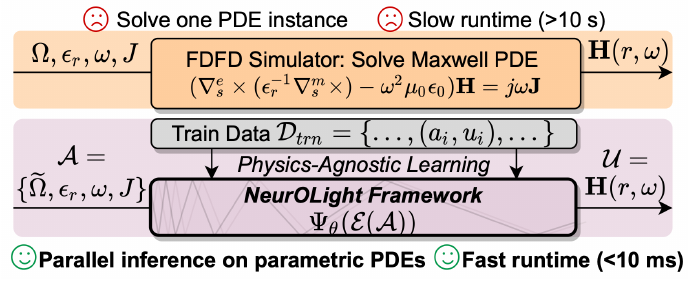}
    \label{fig:neur_b}
    }
    \vspace{-10pt}
    \subfloat[]{\includegraphics[width=0.95\columnwidth]{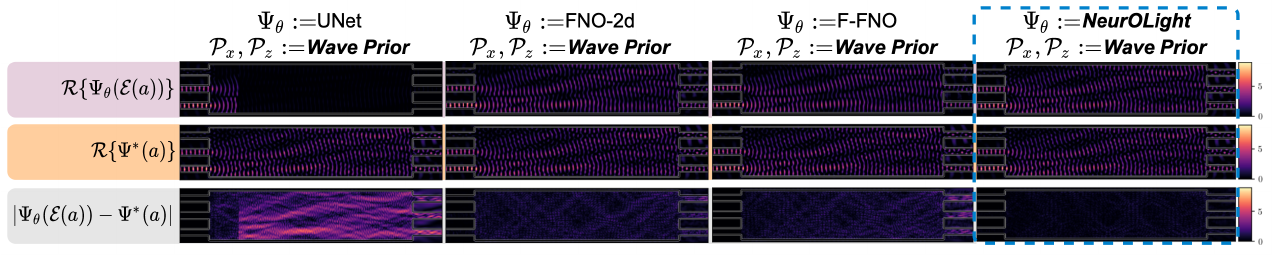}
    \label{fig:neur_c}
    }
    \caption{(a) Runtime comparison of FDFD simulation and \nameneur framework. 
    (b) \nameneur learns a family of parametric Maxwell PDEs for ultra-fast optical field prediction.
    (c) MMI field prediction across different models. The first row shows the real part of the predicted field, the second row shows the FDFD-simulated ground-truth field, and the third row shows the prediction error.
    }
    \label{fig:neurolight}
\end{figure}

%% file: doc/3_3_MAPS.tex
\input{figtex/fig_maps}
\subsection{AI-Infused Inverse Design Framework}
\label{sec:ID}

Achieving the stringent requirements for extreme performance, compact footprint, and energy efficiency in next-generation electronic-photonic AI systems necessitates meticulous device-level design. Conventional photonic design typically relies on extensive physics-based heuristics and iterative trial-and-error, which imposes significant demands on designers. These manual tuning processes are often inefficient and restricted to intuitive design spaces, frequently resulting in larger device footprints. 
In contrast, \textbf{inverse design}~\cite{NP_minkov2020inverse} formulates the design process as an optimization problem guided by well-defined objectives and constraints, substantially reducing the reliance on detailed physics expertise. By enabling the exploration of high-dimensional, non-intuitive design spaces, inverse design allows for the creation of significantly more compact devices than traditional methods can typically achieve.

To accelerate the inverse design process and enable large-scale deployment, researchers have explored the integration of \emph{AI-based surrogate models to replace conventional solvers}, such as those discussed in Section~\ref{sec:solvers}. However, this transition introduces critical methodological challenges: (1) the absence of standardized datasets; (2) the lack of reproducible evaluation metrics for benchmarking AI models; and (3) the frequently overstated effectiveness of existing surrogates. While these models may produce visually plausible forward predictions, they often fail to provide the reliable gradients necessary for optimization, thereby limiting their practical utility in replacing numerical solvers.

To address these challenges, we \textbf{propose \namemaps~\cite{maps}, a modular, open-source infrastructure for AI-augmented photonic simulation and inverse design} 
(Fig.~\ref{fig:maps_a}). This platform unifies data generation, model training, and fabrication-aware optimization into a cohesive framework. Specifically, \namemapsdata provides a flexible data acquisition engine capable of generating multi-fidelity, richly annotated datasets through optimization-aware sampling strategies. \namemapstrain supports customizable training pipelines for both data-driven and physics-informed models, incorporating standardized evaluation metrics such as field prediction accuracy and adjoint gradient alignment. Furthermore, \namemapsinvdes abstracts the complexity of adjoint-based optimization while enabling seamless integration with pretrained neural solvers and differentiable fabrication models~\cite{boson}. 

As a practical demonstration, Fig.~\ref{fig:maps_b} (top) shows an inverse-designed photonic bend optimized with \nameneur, achieving a $\sim$100$\times$ speedup over conventional numerical solvers, while the final design is rigorously validated using high-fidelity FDFD simulations. And the bottom one displays the optimization trajectories for this device, highlighting the convergence characteristics of the neural solver compared to a traditional numerical approach. Collectively, \namemaps facilitates scalable and variation-robust photonic inverse design, establishing a foundation for reproducible AI-for-optics research and real-world deployment.

%% file: figtex/fig_maps.tex
\begin{figure}[t]
  \centering
    \subfloat[]{\includegraphics[width=0.6\columnwidth]{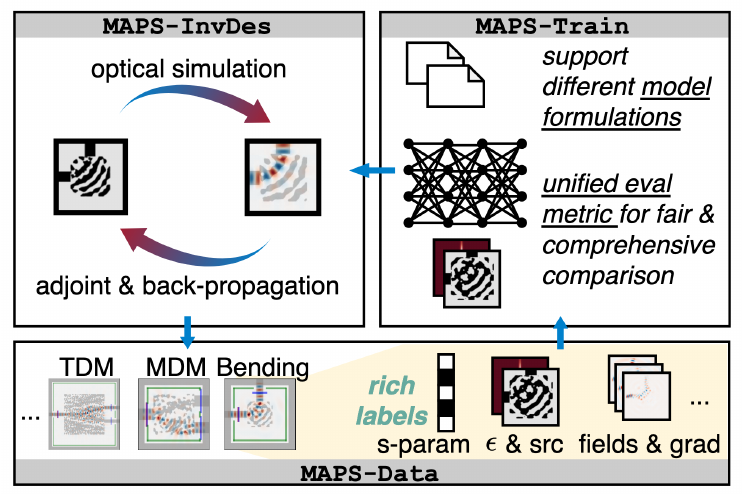}
    \label{fig:maps_a}
    }
    \subfloat[]{\includegraphics[width=0.35\columnwidth]{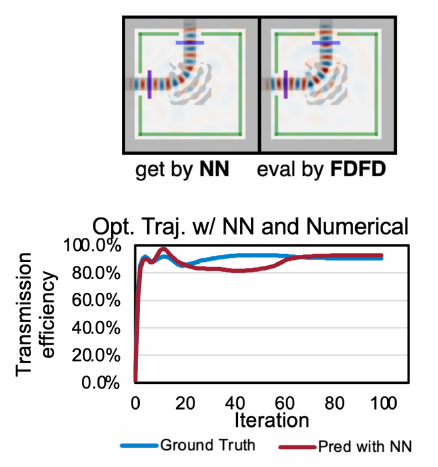}
    \label{fig:maps_b}
    }
  \caption{
  (a) Illustration of our \namemaps~\cite{maps} infrastructure based on three synergistic sub-modules: \namemapsdata, \namemapstrain, and \namemapsinvdes, targeting AI-assisted photonic simulation and inverse design.
  (b) Top: The electrical field of the inverse-designed bend predicted by NN and verified by FDFD; bottom: Optimization trajectory driven by NN-predicted gradients. Transmission efficiency calculated based on NN-predicted fields and FDFD-simulated fields is shown for comparison.
  }
  \label{fig:maps}
\end{figure}

%% file: doc/3_4_Sprint.tex
\input{figtex/fig_sp2rint}
\subsection{Scalable Inverse Training Algorithm of Meta-Optical Neural Networks}
In this section, we demonstrate the extension of our inverse design methodology \textbf{from component to the systematic development} for scalable Meta-Optical Neural Network (MONN) training. 
MONNs are multi-layer Diffractive Optical Neural Networks (DONNs), harnessing engineered metasurfaces to perform high-capacity, energy-efficient analog computation at the speed of light. 
By precisely modulating the phase and amplitude of incident light through arrays of subwavelength meta-atoms, these systems enable complex mathematical operations within high parallelism and near-zero energy cost.

However, training MONNs to achieve \emph{high accuracy while ensuring physical feasibility remains a significant challenge}. Conventional design strategies often rely on the \emph{Local Periodic Approximation (LPA)}~\cite{wu2019neuromorphic}, which oversimplifies metasurfaces as idealized, element-wise phase masks. 
This approach neglects critical inter-element optical interactions, often resulting in designs that are physically unrealizable or suffer from severe performance degradation upon implementation. 
Conversely, \emph{simulation-in-the-loop training} methods embed full-wave electromagnetic simulations directly into the optimization process. 
While accurate, these approaches are prohibitively expensive due to their cubic $O(n^3)$ complexity, which prevents them from scaling to large-capacity or multi-layer systems.

To bridge this gap, we \textbf{integrate inverse design with DONN training in a scalable framework} \textbf{SP$^2$RINT} (Spatially-Decoupled Physics-Inspired ProgRessive INverse OpTimization)~\cite{ma2025sp2rint} that formulates MONN design as a PDE-constrained learning problem. 
SP$^2$RINT introduces several primary innovations to overcome the scalability bottleneck in brute-force inverse design. 
First, \textbf{Spatially Decoupled Simulation} exploits the \emph{natural locality of meta-atom interactions and the field smoothness} inherent in diffraction to partition the metasurface into independently solvable patches. 
As illustrated in Fig.~\ref{fig:sp2rint1}, this approach transforms the computational complexity from \emph{cubic to near-linear}, enabling the parallel simulation of large-scale metasurfaces. 
Second, \textbf{Progressive PDE-Constrained Learning} avoids imposing rigid physical constraints at the onset of training by employing a progressive soft projection strategy. 
Metasurface responses are initially \emph{relaxed into freely-trainable transfer matrices} and then gradually projected onto the Maxwell-constrained subspace via inverse design. 
This prevents the optimization from becoming trapped in poor local optima and ensures the final design is both high-performing and physically implementable.

Our SP$^2$RINT can \emph{inversely optimize} DONNs with physically implementable metasurfaces to realize digital-comparable AI inference accuracy while being \textbf{1825$\times$ faster} than traditional simulation-in-the-loop approaches. 
To visualize how our method can inversely optimize metasurfaces with more accurate modeling than prior approaches, Fig.~\ref{fig:sp2rint2} compares the estimated and simulated optical fields within the diffractive feature extractor during inference. 
This pipeline can be potentially further accelerated by replacing conventional solvers with our AI-based photonic solvers introduced in Section~\ref{sec:solvers} as an AI-accelerated MONN training framework, which is left for our future exploration. 
\input{figtex/fig_sp2rint2}

%% file: figtex/fig_sp2rint.tex
\begin{figure}
    \centering
    \includegraphics[width=\columnwidth]{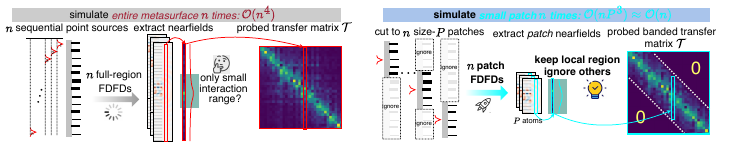}
    \vspace{-5pt}
    \caption{Our proposed spatially-decoupled transfer matrix probing
method SP$^2$RINT~\cite{ma2025sp2rint} cuts the metasurface into small patches for fast simulation that reduces complexity from cubic to linear.
    }
    \label{fig:sp2rint1}
     \vspace{-5pt}
\end{figure}

%% file: figtex/fig_sp2rint2.tex
\begin{figure}
    \centering
    \includegraphics[width=0.9\columnwidth]{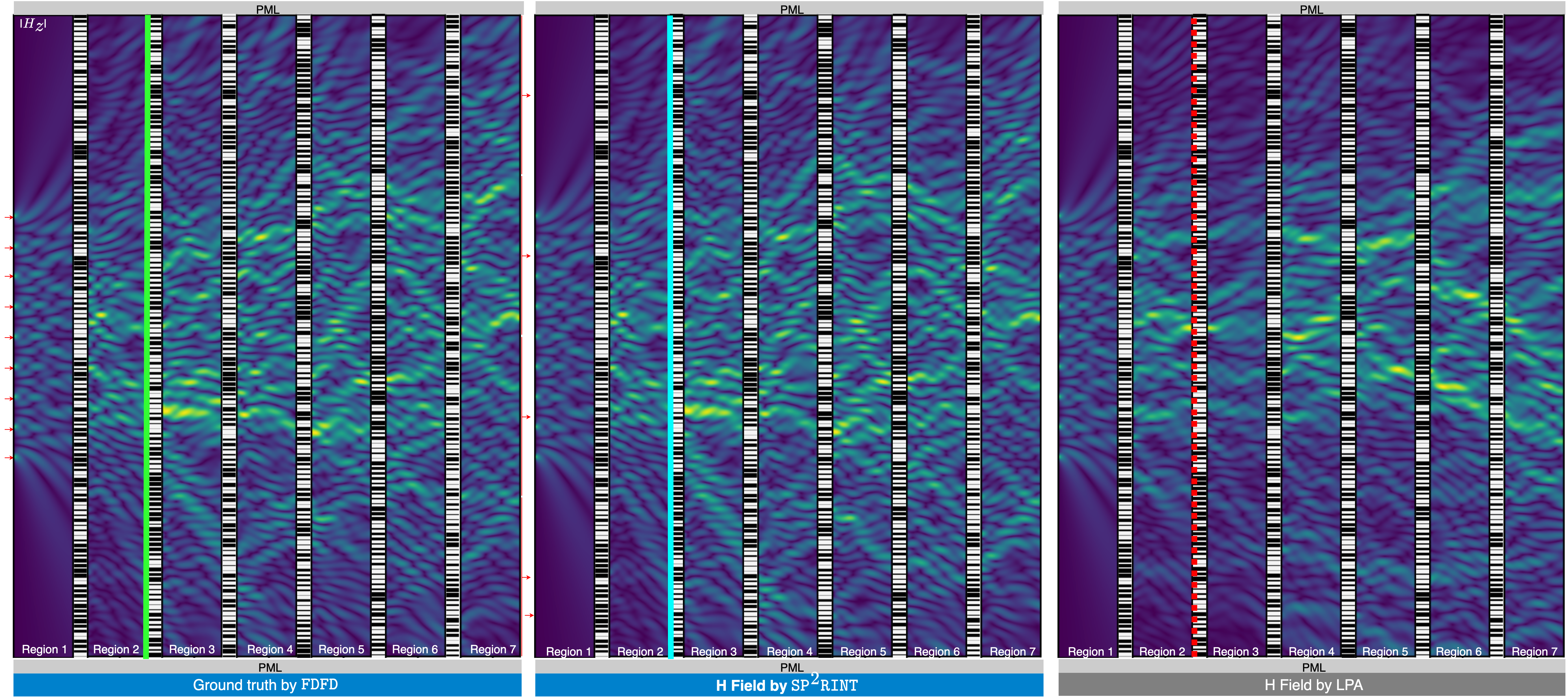}
    \vspace{5pt}
    \caption{\(|H_z|\) field comparison fora 6-layer diffraction system consisting of 128-meta-atoms metasurfaces.
\namesprint captures the transfer matrix accurately and outperforms LPA,
closely matching the FDFD result.
    }
    \label{fig:sp2rint2}
\end{figure}

%% file: doc/5_Conclusion.tex
\section{Conclusion and Outlook}
\label{sec:discussion}
The developments surveyed here point toward an emerging EPDA stack that leverages AI to accelerate, stabilize, and systematize electronic-photonic AI system design. 
Our system simulation framework SimPhony enables fast cross-layer evaluation by bridging device-level photonic models with circuit and architecture metrics, using compact layout-aware models.
Then, our proposed physical AI-based Maxwell solvers further reduce the cost of electromagnetic simulation by orders of magnitude. 
Based on these simulators, we present an open-source infrastructure for AI-augmented photonic simulation and inverse design that unifies data generation, model training, and fabrication-aware optimization in a cohesive framework.
The frameworks of SP$^{2}$RINT extend this inverse design idea to meta-optical neural networks, demonstrating that PDE-constrained training can be made scalable by combining relaxed representations, inverse design, and spatial decomposition.

Looking forward, automating electronic-photonic design will be pivotal to unlocking the next wave of advances in computing and communications. Realizing the full potential of EPIC AI platforms calls for a holistic photonic-electronic co-design framework that jointly optimizes devices, circuits, and algorithms. 
In particular, architectural exploration should couple system-level evaluation with device-level parameter optimization and sensitivity analysis, so that physical variations can be directly mapped to end-to-end AI metrics. Beyond optimizing individual components, a key frontier is \emph{system-scale} inverse design: embedding differentiable inverse-design engines into EPDA can automatically synthesize workload-specific photonic primitives and feed compact, fabrication-aware models back into the co-design loop. Equally important is rigorous \emph{circuit-level} simulation to capture dominant electronic overheads and their interactions with active photonics. Ultimately, EPDA should enable ``push-button'' arch-to-layout automation, translating high-level specifications into netlists and verified layouts via automated mapping, placement, and routing, layout-aware parasitic extraction, and multi-physics verification, thereby reducing iteration time and lowering the barrier to entry. With a more intelligent and accessible design ecosystem, integrated photonics can be rapidly translated into practical solutions for demanding real-world applications, sustaining innovation across the photonic computing stack.

%% file: main.bbl
\begin{thebibliography}{10}

\bibitem{ning2024photonic}
Ning, S., Zhu, H., Feng, C., Gu, J., Jiang, Z., Ying, Z., Midkiff, J., Jain, S., Hlaing, M.~H., Pan, D.~Z., et~al., ``Photonic-electronic integrated circuits for high-performance computing and ai accelerators,'' {\em Journal of Lightwave Technology}  (2024).

\bibitem{NP_Nature_ahmed}
Ahmed, S.~R., Baghdadi, R., Bernadskiy, M., Bowman, N., Braid, R., Carr, J., Chen, C., Ciccarella, P., Cole, M., Cooke, J., et~al., ``Universal photonic artificial intelligence acceleration,'' {\em Nature}~{\bf 640}(8058),  368--374 (2025).

\bibitem{NP_Nature2021_Xu}
Xu, X., Tan, M., Corcoran, B., Wu, J., Boes, A., Nguyen, T.~G., Chu, S.~T., Little, B.~E., Hicks, D.~G., Morandotti, R., Mitchell, A., and Moss, D.~J., ``{11 TOPS photonic convolutional accelerator for optical neural networks},'' {\em Nature}  (2021).

\bibitem{NP_Nature2021_Feldmann}
Feldmann, J., Youngblood, N., Karpov, M., Gehring, H., Li, X., Stappers, M., Gallo, M.~L., Fu, X., Lukashchuk, A., Raja, A., Liu, J., Wright, D., Sebastian, A., Kippenberg, T., Pernice, W., and Bhaskaran, H., ``Parallel convolutional processing using an integrated photonic tensor core,'' {\em Nature}  (2021).

\bibitem{Afifi2025ASTRA}
Afifi, S., Alo, O., Thakkar, I., and Pasricha, S., ``Astra: A stochastic transformer neural network accelerator with silicon photonics,'' {\em ACM Transactions on Embedded Computing Systems}  (2025).

\bibitem{tang2025optical}
Tang, Y., Chen, R., Lou, M., Fan, J., Yu, C., Nonaka, A., Yao, Z., and Gao, W., ``Optical neural engine for solving scientific partial differential equations,'' {\em Nature Communications}~{\bf 16}(1),  4603 (2025).

\bibitem{Najafi2025OptoAligner}
Najafi, D., Errahmouni~Barkam, H., Ghanaatian, Z., Morsali, M., Chen, H., Das, T., Roohi, A., Mercati, P., Nikdast, M., Imani, M., and Angizi, S., ``Opto-aligner: Optical near-sensor architecture for accelerating {DNA} pre-alignment filtering,'' {\em IEEE Journal on Emerging and Selected Topics in Circuits and Systems}  (2025).

\bibitem{dong2024partial}
Dong, B., Br{\"u}ckerhoff-Pl{\"u}ckelmann, F., Meyer, L., Dijkstra, J., Bente, I., Wendland, D., Varri, A., Aggarwal, S., Farmakidis, N., Wang, M., et~al., ``Partial coherence enhances parallelized photonic computing,'' {\em Nature}~{\bf 632}(8023),  55--62 (2024).

\bibitem{Hua2025NaturePhotonicAccelerator}
Hua, S., Divita, E., Yu, S., Peng, B., et~al., ``An integrated large-scale photonic accelerator with ultralow latency,'' {\em Nature}~{\bf 640},  361--367 (2025).

\bibitem{hua2025integrated}
Hua, S., Divita, E., Yu, S., Peng, B., Roques-Carmes, C., Su, Z., Chen, Z., Bai, Y., Zou, J., Zhu, Y., et~al., ``An integrated large-scale photonic accelerator with ultralow latency,'' {\em Nature}~{\bf 640}(8058),  361--367 (2025).

\bibitem{zhou2025toward}
Zhou, H., Ma, P., and Gu, J., ``Toward intelligent electronic-photonic design automation for large-scale photonic integrated circuits: from device inverse design to physical layout generation,'' in [{\em Optical Design Automation}{\nolinebreak\hspace{0.1em}]},   {\bf 13601},  69--78, SPIE (2025).

\bibitem{bogaerts2018silicon}
Bogaerts, W. and Chrostowski, L., ``Silicon photonics circuit design: methods, tools and challenges,'' {\em Laser \& Photonics Reviews}~{\bf 12}(4),  1700237 (2018).

\bibitem{yin2025simphony}
Yin, Z., Zhang, M., Gangi, N., Huang, R., Zhang, J., and Gu, J., ``Simphony: A device-circuit-architecture cross-layer modeling and simulation framework for heterogeneous electronic-photonic ai system,'' in [{\em 2025 62nd ACM/IEEE Design Automation Conference (DAC)}{\nolinebreak\hspace{0.1em}]},   1--7, IEEE (2025).

\bibitem{gu2022NeurOLight}
Gu, J., Gao, Z., Feng, C., Zhu, H., Chen, R.~T., Boning, D.~S., and Pan, D.~Z., ``Neurolight: A physics-agnostic neural operator enabling parametric photonic device simulation,'' in [{\em Conference on Neural Information Processing Systems (NeurIPS)}{\nolinebreak\hspace{0.1em}]},  (2022).

\bibitem{zhu2024Pace}
Zhu, H., Cong, W., Chen, G., Ning, S., Chen, R., Gu, J., and Pan, D.~Z., ``Pace: Pacing operator learning to accurate optical field simulation for complicated photonic devices,'' in [{\em Conference on Neural Information Processing Systems (NeurIPS)}{\nolinebreak\hspace{0.1em}]},  (2024).

\bibitem{pcma2024pic2osim}
Ma, P., Yang, H., Gao, Z., Boning, D.~S., and Gu, J., ``{PIC$^2$O-Sim: A physics-inspired causality-aware dynamic convolutional neural operator for ultra-fast photonic device time-domain simulation},'' {\em APL Photonics}~{\bf 10},  036104 (03 2025).

\bibitem{maps}
Mal, P., Gao, Z., Zhang, M., Yang, H., Ren, M., Huang, R., Boning, D.~S., and Gu, J., ``{MAPS: Multi-Fidelity AI-Augmented Photonic Simulation and Inverse Design Infrastructure},'' in [{\em 2025 Design, Automation \& Test in Europe Conference (DATE)}{\nolinebreak\hspace{0.1em}]},   1--6 (2025).

\bibitem{ma2025sp2rint}
Ma, P., Yin, Z., Jing, Q., Gao, Z., Gangi, N., Zhang, B., Huang, T.-W., Huang, Z., Boning, D.~S., Yao, Y., et~al., ``Sp2rint: Spatially-decoupled physics-inspired progressive inverse optimization for scalable, pde-constrained meta-optical neural network training,'' {\em arXiv preprint arXiv:2505.18377}  (2025).

\bibitem{NP_zhu2024lightening}
Zhu, H., Gu, J., Wang, H., Jiang, Z., Zhang, Z., Tang, R., Feng, C., Han, S., Chen, R.~T., and Pan, D.~Z., ``Lightening-transformer: A dynamically-operated optically-interconnected photonic transformer accelerator,'' in [{\em 2024 IEEE International Symposium on High-Performance Computer Architecture (HPCA)}{\nolinebreak\hspace{0.1em}]},   686--703, IEEE (2024).

\bibitem{NP_zhang2024tempo}
Zhang, M., Yin, D., Gangi, N., Begovi{\'c}, A., Chen, A., Huang, Z.~R., and Gu, J., ``Tempo: Efficient time-multiplexed dynamic photonic tensor core for edge ai with compact slow-light electro-optic modulator,'' {\em Journal of Applied Physics}~{\bf 135}(22) (2024).

\bibitem{NP_ICCAD24_Gu}
Yin, Z., Gangi, N., Zhang, M., Zhang, J., Huang, R., and Gu, J.,  [{\em SCATTER: Algorithm-Circuit Co-Sparse Photonic Accelerator with Thermal-Tolerant, Power-Efficient In-situ Light Redistribution}{\nolinebreak\hspace{0.1em}]}, Association for Computing Machinery, New York, NY, USA (2025).

\bibitem{Fan2025DeepLearningPhotonicCAD}
Fan, J.~A., ``Deep learning approaches to photonic computer aided design,'' in [{\em AI and Optical Data Sciences VI}{\nolinebreak\hspace{0.1em}]},   PC1337512, SPIE (2025).

\bibitem{boson}
Ma, P., Gao, Z., Begovic, A., Zhang, M., Yang, H., Ren, H., Huang, R., Boning, D.~S., and Gu, J., ``{BOSON$^{-1}$: Understanding and Enabling Physically-Robust Photonic Inverse Design with Adaptive Variation-Aware Subspace Optimization},'' in [{\em 2025 Design, Automation \& Test in Europe Conference (DATE)}{\nolinebreak\hspace{0.1em}]},  (2025).

\bibitem{chen2022wavey}
Chen, M., Lupoiu, R., Mao, C., Huang, D.-H., Jiang, J., Lalanne, P., and Fan, J.~A., ``Wavey-net: physics-augmented deep-learning for high-speed electromagnetic simulation and optimization,'' in [{\em High Contrast Metastructures XI}{\nolinebreak\hspace{0.1em}]},   {\bf 12011},  63--66, SPIE (2022).

\bibitem{lu2019deeponet}
Lu, L., Jin, P., Pang, G., Zhang, Z., and Karniadakis, G.~E., ``Learning nonlinear operators via {DeepONet} based on the universal approximation theorem of operators,'' {\em Nature Machine Intelligence}~{\bf 3}(3),  218--229 (2021).

\bibitem{li2020fourier}
Li, Z., Kovachki, N., Azizzadenesheli, K., Liu, B., Bhattacharya, K., Stuart, A., and Anandkumar, A., ``Fourier neural operator for parametric partial differential equations,'' in [{\em Proc.~ICLR}{\nolinebreak\hspace{0.1em}]},  (2021).

\bibitem{Su2025SuperPhysNet}
Su, Y., Chen, H., Dai, G., Ma, Y., and Tong, Y., ``Superphys-net: A physics-informed super-resolution electromagnetic simulator for nanophotonic devices,'' in [{\em Proc.~DATE}{\nolinebreak\hspace{0.1em}]},  (2025).

\bibitem{oskooi2010meep}
Oskooi, A.~F., Roundy, D., Ibanescu, M., Bermel, P., Joannopoulos, J.~D., and Johnson, S.~G., ``Meep: A flexible free-software package for electromagnetic simulations by the fdtd method,'' {\em Computer Physics Communications}~{\bf 181}(3),  687--702 (2010).

\bibitem{NP_minkov2020inverse}
Minkov, M., Williamson, I.~A., Andreani, L.~C., Gerace, D., Lou, B., Song, A.~Y., Hughes, T.~W., and Fan, S., ``Inverse design of photonic crystals through automatic differentiation,'' {\em Acs Photonics}~{\bf 7}(7),  1729--1741 (2020).

\bibitem{wu2019neuromorphic}
Wu, Z., Zhou, M., Khoram, E., Liu, B., and Yu, Z., ``Neuromorphic metasurface,'' {\em Photonics Research}~{\bf 8}(1),  46--50 (2019).

\end{thebibliography}
